\documentclass[10pt]{article}
\usepackage{graphicx}

\def\Title#1{\begin{center} {\Large #1 } \end{center}}
\def\Author#1{\begin{center}{ \sc #1} \end{center}}
\def\Address#1{\begin{center}{ \it #1} \end{center}}

\newcommand\pubblock{\rightline{\begin{tabular}{l} Proceedings of the Fifth Annual LHCP\\ \pubnumber\\
         \pubdate  \end{tabular}}}

\newenvironment{Abstract}{\begin{quotation} \begin{center} 
             \large ABSTRACT \end{center}\bigskip 
      \begin{center}\begin{large}}{\end{large}\end{center} \end{quotation}}

\newenvironment{Presented}{\begin{quotation} \begin{center} 
             PRESENTED AT\end{center}\bigskip 
      \begin{center}\begin{large}}{\end{large}\end{center} \end{quotation}}

\def\Acknowledgements{\bigskip  \bigskip \begin{center} \begin{large}
             \bf ACKNOWLEDGEMENTS \end{large}\end{center}}

\def\beq{\begin{equation}}
\def\eeq#1{\label{#1}\end{equation}}
\def\eeqn{\end{equation}}

\def\beqa{\begin{eqnarray}}
\def\eeqa#1{\label{#1}\end{eqnarray}}
\def\eeqan{\end{eqnarray}}

\let\bar=\overbar

\def\Dslash{\not{\hbox{\kern-4pt $D$}}}
\def\dslash{\not{\hbox{\kern-2pt $\del$}}}

\def\msb{{\bar{\ssstyle M \kern -1pt S}}}

\usepackage{color}

\newcommand\pubnumber{ ATL-PHYS-PROC-2017-XXX }

\newcommand\pubdate{\today}

\def\affiliation{
On behalf of the CMS Experiment, \\
Department of Physics \\
Northeastern University, Boston, USA}

\begin{document}

\large
\begin{titlepage}
\pubblock

\vfill
\Title{  Latest results on di-Higgs boson production with CMS  }
\vfill

\Author{ David Morse  }
\Address{\affiliation}
\vfill
\begin{Abstract}

The latest results on searches for production of two Higgs bosons with the CMS detector using the 2016 CERN LHC dataset are presented.
\end{Abstract}
\vfill

\begin{Presented}
The Fifth Annual Conference\\
 on Large Hadron Collider Physics \\
Shanghai Jiao Tong University, Shanghai, China\\ 
May 15-20, 2017
\end{Presented}
\vfill
\end{titlepage}
\def\thefootnote{\fnsymbol{footnote}}
\setcounter{footnote}{0}
%

\normalsize 


\section{Introduction}
Di-higgs boson production provides a crucial test of standard model (SM) electroweak symmetry breaking as well as a search for new physics beyond the standard model (BSM). The CMS experiment~\cite{Chatrchyan:2008aa} at the CERN LHC has a suite of di-higgs searches which are disjoint and complementary across a large mass range and parameter space.

\begin{figure}[htb]
\centering
\hspace*{-2.5cm}\raisebox{6mm}[0pt][0pt]{\includegraphics[width=.7\textwidth]{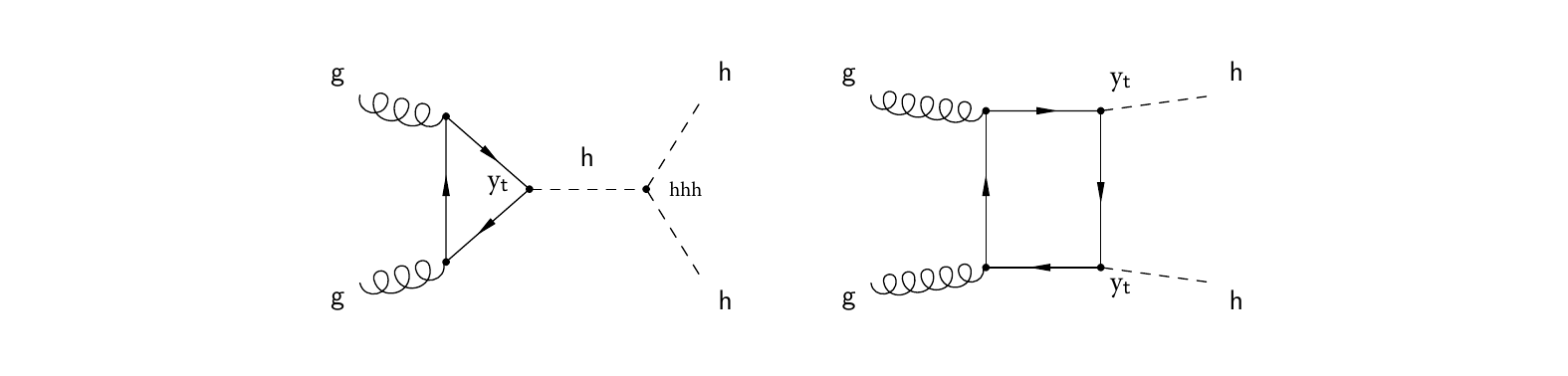}}
\hspace*{-.5cm}\includegraphics[width=0.25\textwidth]{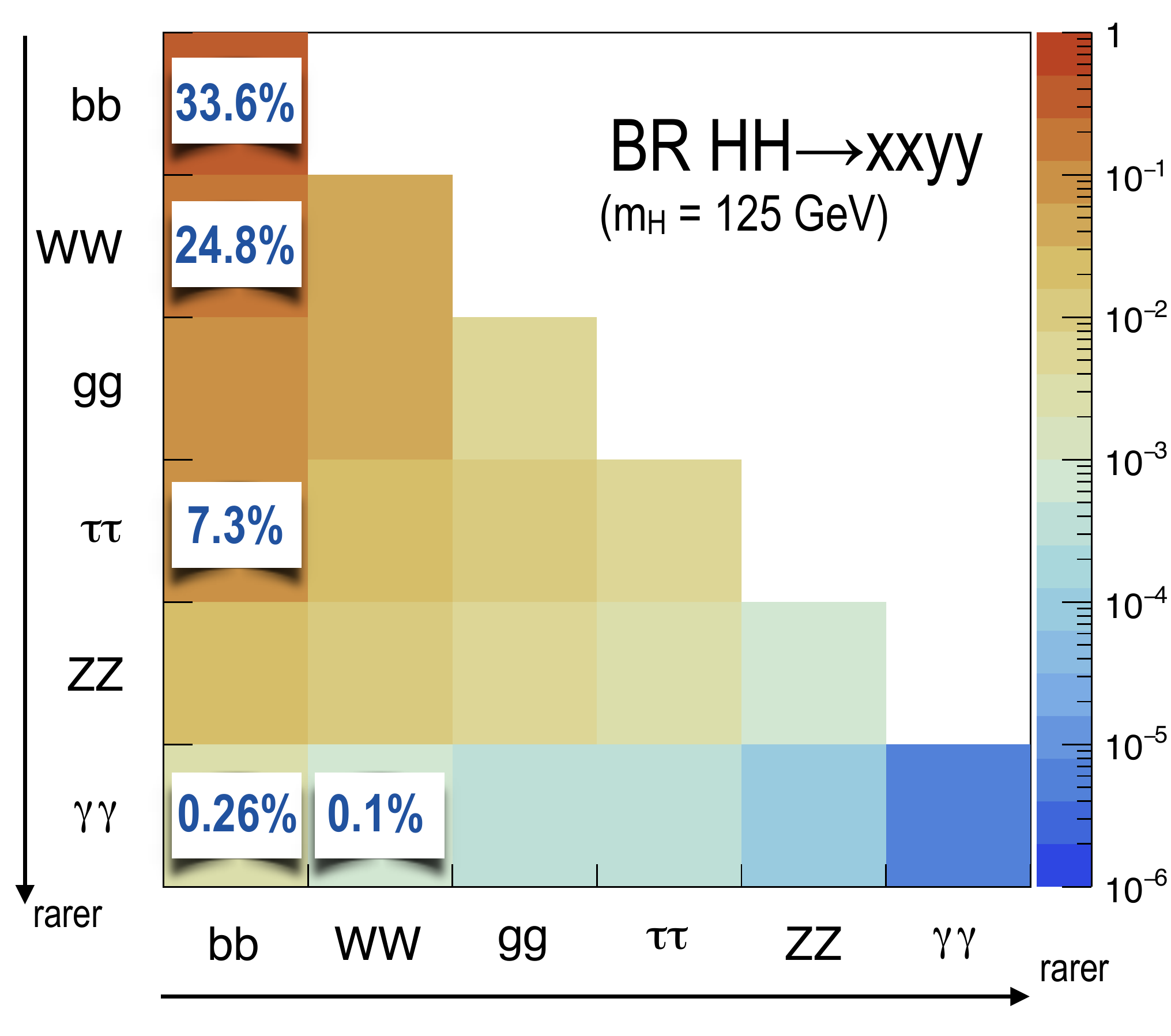}
\caption{Dominant di-Higgs boson production diagrams via gluon-gluon fusion (left), possible di-Higgs boson final states (right).}
\label{fig:feynman}
\end{figure}

Di-Higgs boson production can be categorized into resonant and non-resonant categories. SM di-Higgs boson production is dominated by non-resonant gluon-gluon fusion production, as shown in Figure \ref{fig:feynman} (left).

\noindent SM di-Higgs production provides insight into the nature of electroweak symmetry breaking through access to the top-Higgs Yukawa coupling $y_t$ and the Higgs boson trilinear coupling $\lambda_{hhh}$, seen here in the Higgs field potential:

\[ V=\frac{m_{h}^2}{2}h^2+\lambda_{hhh}vh^3+\frac{\lambda_{hhhh}}{4}h^4,\quad \lambda_{hhh}=m_{h}^2/(2v^2) \] 

BSM models can lead to non-resonant and resonant di-Higgs production. Non-resonant production would arise from anomalous couplings $\kappa_t=y_{t}'/y_{t}^{SM}, \ \kappa_\lambda=\lambda_{hhh}'/\lambda_{hhh}^{SM}$ and up to 4 new contact interactions which can lead to large modifications in production cross section and kinematic shapes. Resonant production proceeds from BSM models where a new particle $X$ decays to two Higgs bosons $X\rightarrow\mathrm{HH}$.

The numerous decay possibilities of the Higgs boson leads to many di-higgs final state possibilities. All the possible di-Higgs boson final states, along with the branching fractions to some final states are shown in Figure \ref{fig:feynman} (right). Analyses are performed in final states which prioritize high branching fractions and clean final states. All searches in CMS all have one Higgs boson decaying to two b jets, while the decay products of the second Higgs boson vary. The final states considered are H($\mathrm{b\overline{b}}$)H($\mathrm{b\overline{b}}$), H($\mathrm{b\overline{b}}$)H($\gamma\gamma$), H($\mathrm{b\overline{b}}$)H($\ell\nu\ell\nu$), and H($\mathrm{b\overline{b}}$)H($\tau\tau$).

\section{CMS Searches}

\subsection[H(bb)H(bb)]{H($\mathrm{b\overline{b}}$)H($\mathrm{b\overline{b}}$)}

Di-higgs boson searches in the four b jet final state are performed for both resonant~\cite{CMS:2016tlj, CMS:2017gxe} and non-resonant~\cite{CMS:2016foy} production with 2.3--35.9 fb$^{-1}$. This final state has the benefit of the highest di-Higgs boson branching fraction, paired with the challenge of a large multijet background. The resonant analysis is separated into two categories: a 4-jet resolved search~\cite{CMS:2016tlj} and a 2-jet high mass boosted topology~\cite{CMS:2017gxe}.

\begin{figure}[htb]
\centering
\includegraphics[width=0.35\textwidth]{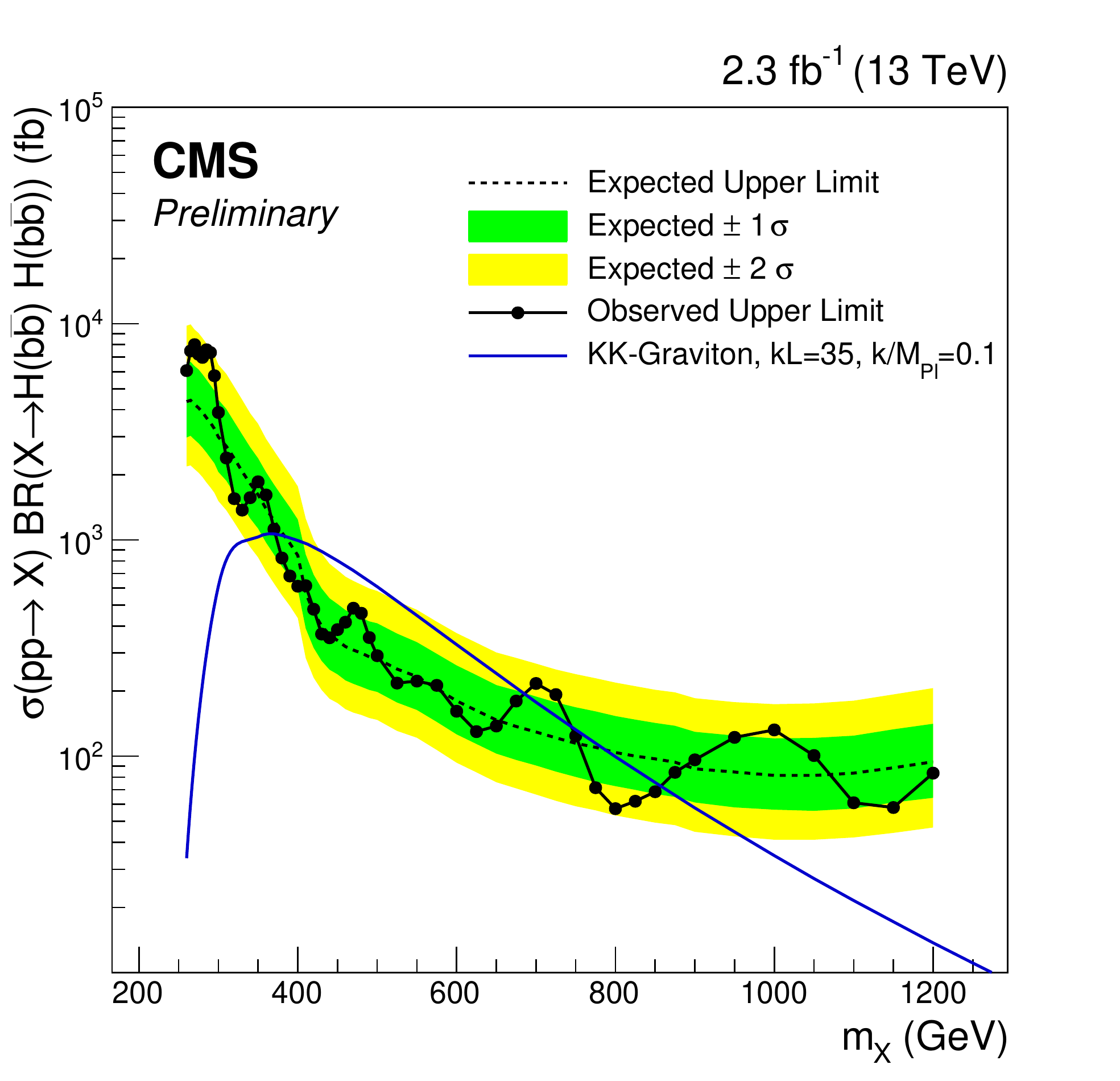}
\caption{Observed and expected upper limits on the cross section times branching fraction for a spin-2 resonance $X\rightarrow$H(b$\mathrm{\overline{b}}$)H(b$\mathrm{\overline{b})}$ at a 95\% confidence level using data corresponding to an integrated luminosity of 2.3$\mathrm{fb}^{-1}$~at $\sqrt{s}$=13 TeV~\cite{CMS:2016tlj}. Theoretical cross sections for the RS1 KK-Graviton decaying to four b jets via Higgs bosons are overlaid.}
\label{fig:4bresolved}
\end{figure}

In the resolved search, a data-driven technique is used to estimate the multijet background. The 95\% confidence level (CL) limits in the resolved search can be seen in Figure \ref{fig:4bresolved}.

\begin{figure}[htb]
\centering
\includegraphics[width=0.35\textwidth]{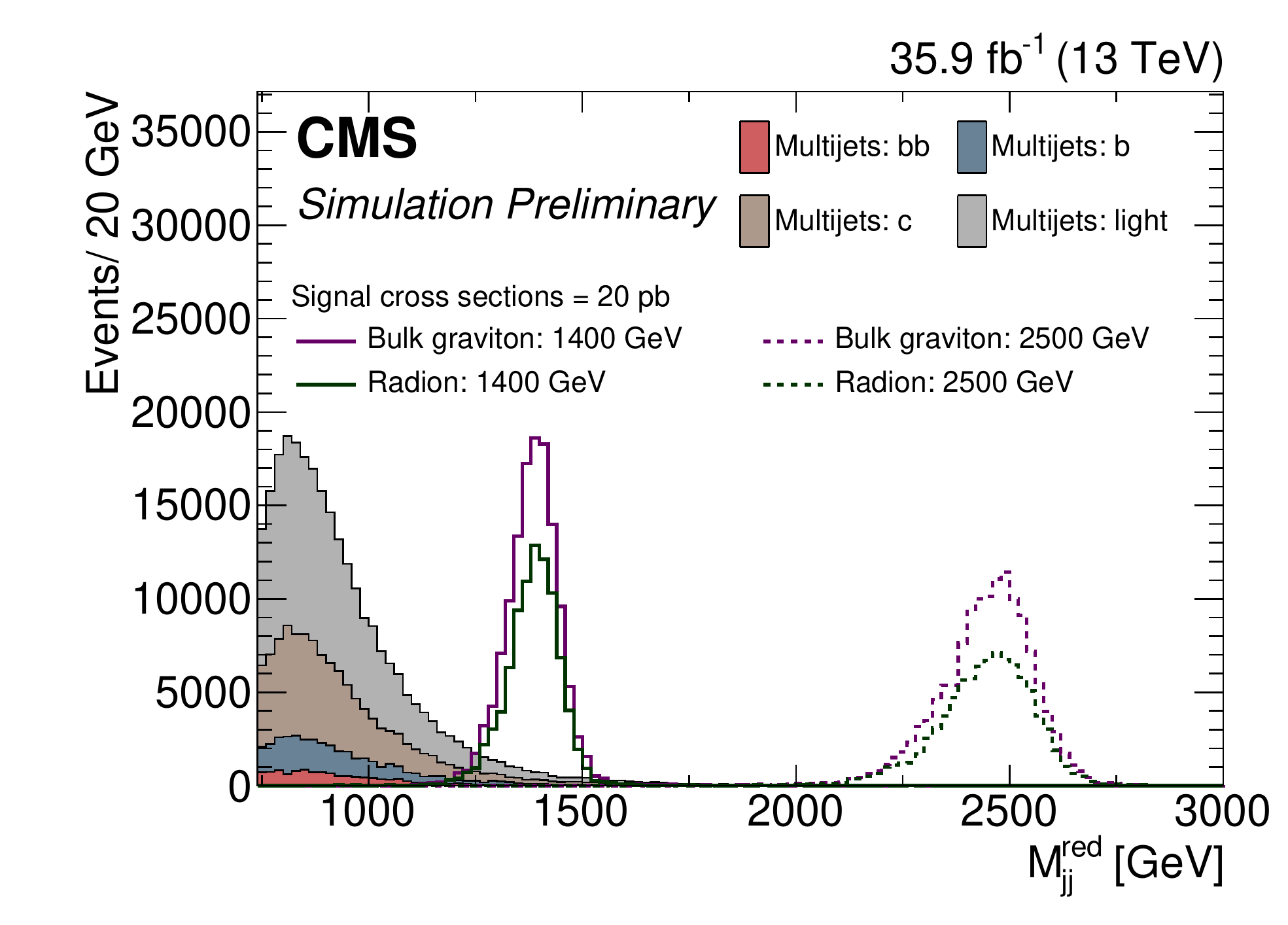}
\includegraphics[width=0.35\textwidth]{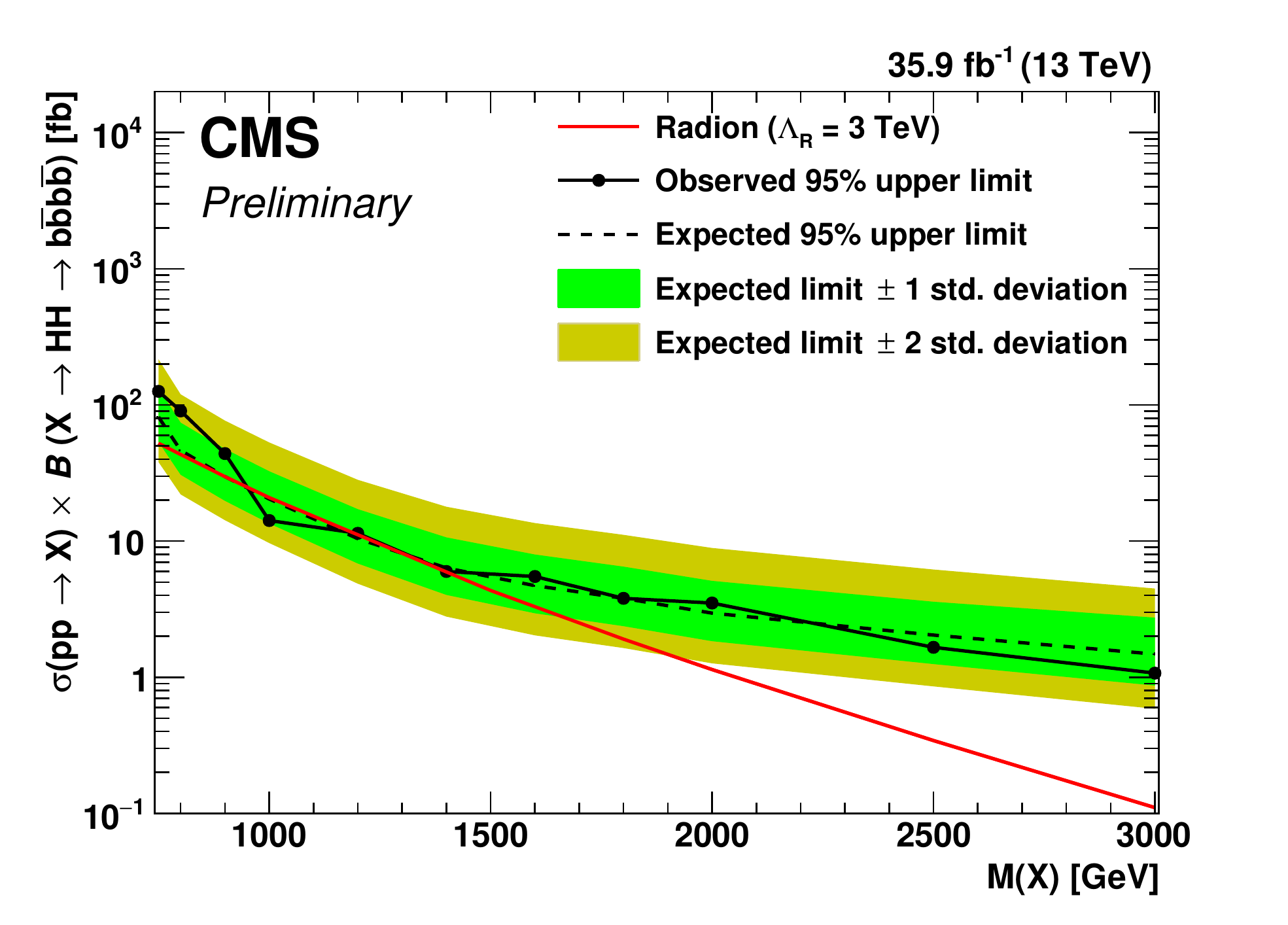}
\caption{Resonant H($\mathrm{b\overline{b}}$)H($\mathrm{b\overline{b}}$) results~\cite{CMS:2017gxe}. Reduced dijet invariant mass $\mathrm{M}_{\mathrm{jj}}^{\mathrm{red}}$ (left). The multijet background components for the different jet flavours are shown, along with a bulk graviton signal of masses 1400, 1800, and 2500 GeV. 95\% CL limits for the spin-0 radion model (right).}
\label{fig:4b_res}
\end{figure}

The high-mass boosted search reconstructs each $\mathrm{H(b\overline{b})}$ system as a single high $p_{T}$ jet, with $105<M_j<130$. Jet substructure techniques and a double b-tagger MVA suppress backgrounds. The multijet background is estimated using data sidebands in the double b-tagger and the invariant mass of the leading jet. Signal to background separation is enhanced by replacing the dijet system invariant mass with the reduced mass $\mathrm{M}_{\mathrm{jj}}^{\mathrm{red}}$: \[ \mathrm{M}_{\mathrm{jj}}^{\mathrm{red}} = \mathrm{M}_{\mathrm{jj}} - (\mathrm{M}_{\mathrm{j1}}-\mathrm{M}_{\mathrm{H}}) - (\mathrm{M}_{\mathrm{j2}}-\mathrm{M}_{\mathrm{H}})\]
\noindent $\mathrm{M}_{\mathrm{jj}}^{\mathrm{red}}$ increases the resolution of the resonance peak, and can be seen in Figure \ref{fig:4b_res} (left). The 95\% CL observed and expected limits are shown in Figure \ref{fig:4b_res} (right), overlaid with the theoretical prediction for a spin-0 radion model.

The non-resonant analysis similarly utilizes a data-driven technique to estimate the multijet background, and the 95\% CL results are summarized in Table \ref{tab:4bResults}.

\begin{table}[ht]
\begin{center}
\begin{small}
\begin{tabular}{l|cccccc}  
Category &  Observed &  Expected &  -2 SD & -1 SD & +1 SD  & +2 SD  \\ \hline
SM H(b$\mathrm{\overline{b}}$)H(b$\mathrm{\overline{b}}$)  &   3880     &     3490      &     2140 & 2540 & 5350 & 8350  \\
\hline
\end{tabular}
\caption{Non-resonant H($\mathrm{b\overline{b}}$)H($\mathrm{b\overline{b}}$) observed and expected limits, in fb. $\pm$1 and 2 standard deviation (SD) limits are also shown for the observed limits}
\label{tab:4bResults}
\end{small}
\end{center}
\end{table}

\subsection[H(bb)H(gamma gamma)]{H($\mathrm{b\overline{b}}$)H($\gamma\gamma$)}

H($\mathrm{b\overline{b}}$)H($\gamma\gamma$) takes advantage of the high H($\mathrm{b\overline{b}}$) branching fraction and the excellent mass resolution of H($\gamma\gamma$). Resonant and non-resonant searches are performed with 2.7 fb$^{-1}$~\cite{CMS:2016vpz}.

A mixed b tag categorization is utilized to produce medium and high purity categories, increasing sensitivity. Signal resolution is increased by replacing the resonance mass $\mathrm{M}_{\mathrm{jj\gamma\gamma}}$ with the modified 4-body mass $\tilde{\mathrm{M}}_{X}$:
\[ \tilde{\mathrm{M}}_{X} = \mathrm{M}_{\mathrm{jj\gamma\gamma}} - \mathrm{M}_\mathrm{jj}+125\ \mathrm{ GeV} \]

\begin{figure}[htb]
\centering
\includegraphics[width=0.35\textwidth]{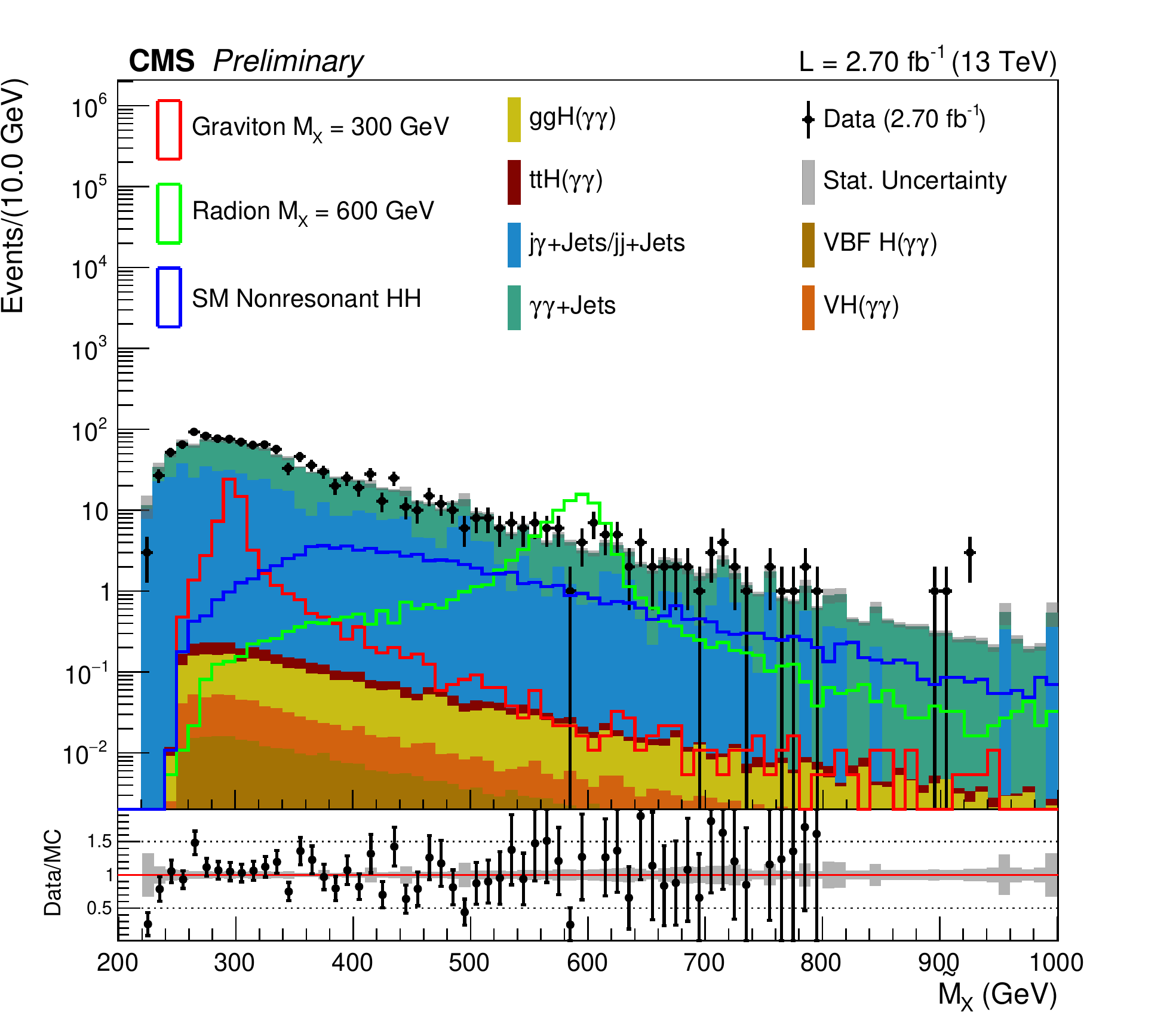}
\raisebox{1mm}[0pt][0pt]{\includegraphics[width=0.4\textwidth]{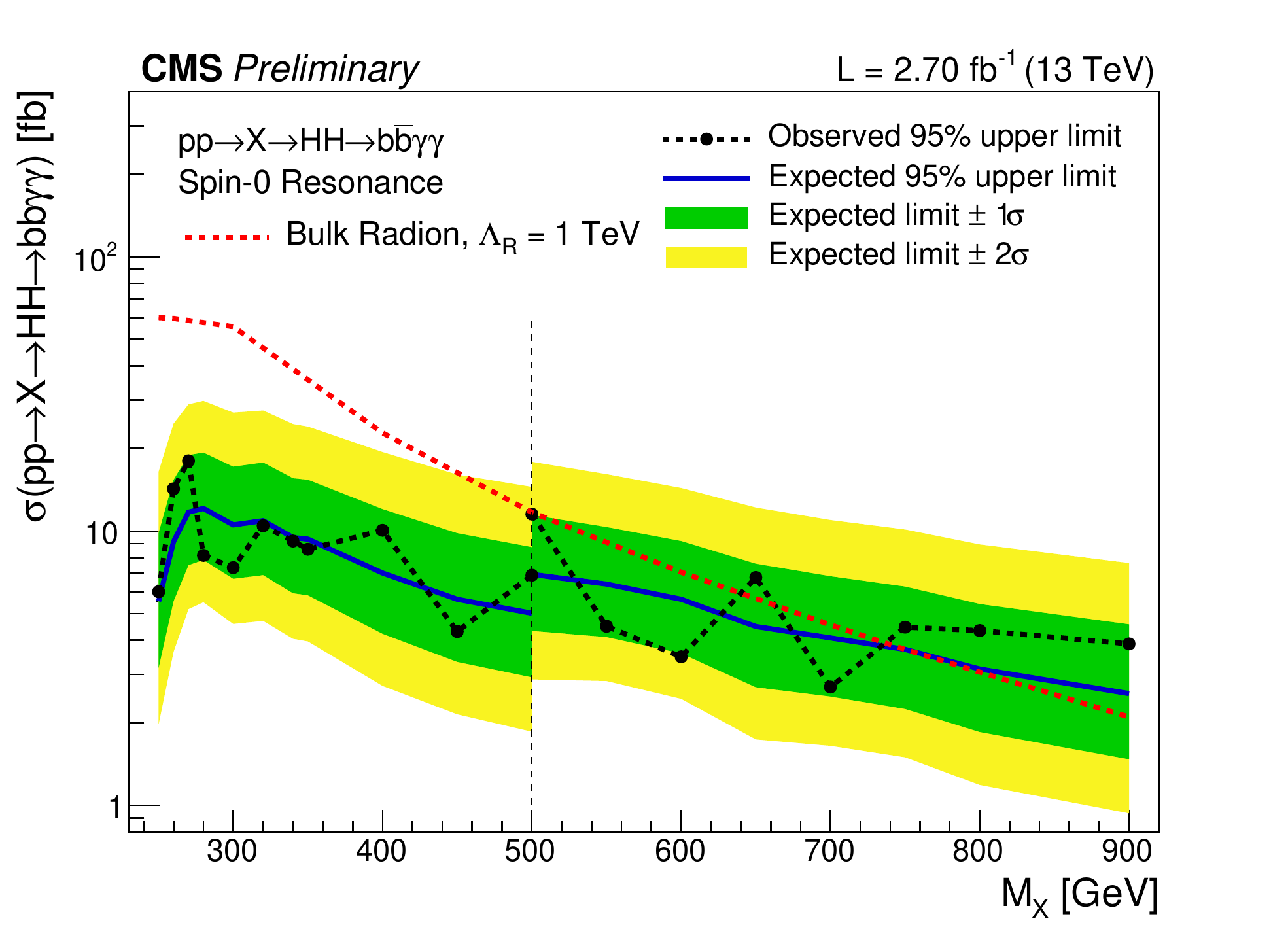}}
\caption{Resonant H($\mathrm{b\overline{b}}$)H($\gamma\gamma$) results~\cite{CMS:2016vpz}. Modified 4-body mass $\tilde{\mathrm{M}}_{X}$ (left); 95\% CL observed and expected limits (right), overlaid with a spin-0 radion signal hypothesis.}
\label{fig:bbgg}
\end{figure}

\noindent The $\tilde{\mathrm{M}}_{X}$ distribution in data, background and signal is shown in Figure \ref{fig:bbgg} (left). The signal and background are modeled in a 2-dimensional space using the dijet and diphoton invariant masses. The signal is modeled using a Gaussian convoluted with a Crystal Ball function, while the background is modeled using Bernstein polynomials.

Limits on the cross section of SM non-resonant H($\mathrm{b\overline{b}}$)H($\gamma\gamma$) production at 95\% CL are observed (expected) at 7.90(7.85) fb, corresponding to 91 (90) times the SM prediction.

\subsection[H(bb)H(l nu l nu)]{H($\mathrm{b\overline{b}}$)H($\ell\nu\ell\nu$)}

H($\mathrm{b\overline{b}}$)H($\ell\nu\ell\nu$) includes decays from both H($\mathrm{b\overline{b}}$)H(WW) and H($\mathrm{b\overline{b}}$)H(ZZ) in final states with two leptons and 2 neutrinos, split into categories of lepton flavor and sign. The search is performed for both resonant and non-resonant production with 35.9 fb$^{-1}$~\cite{CMS:2017ums}.

\begin{figure}[htb]
\centering
\includegraphics[width=.7\textwidth]{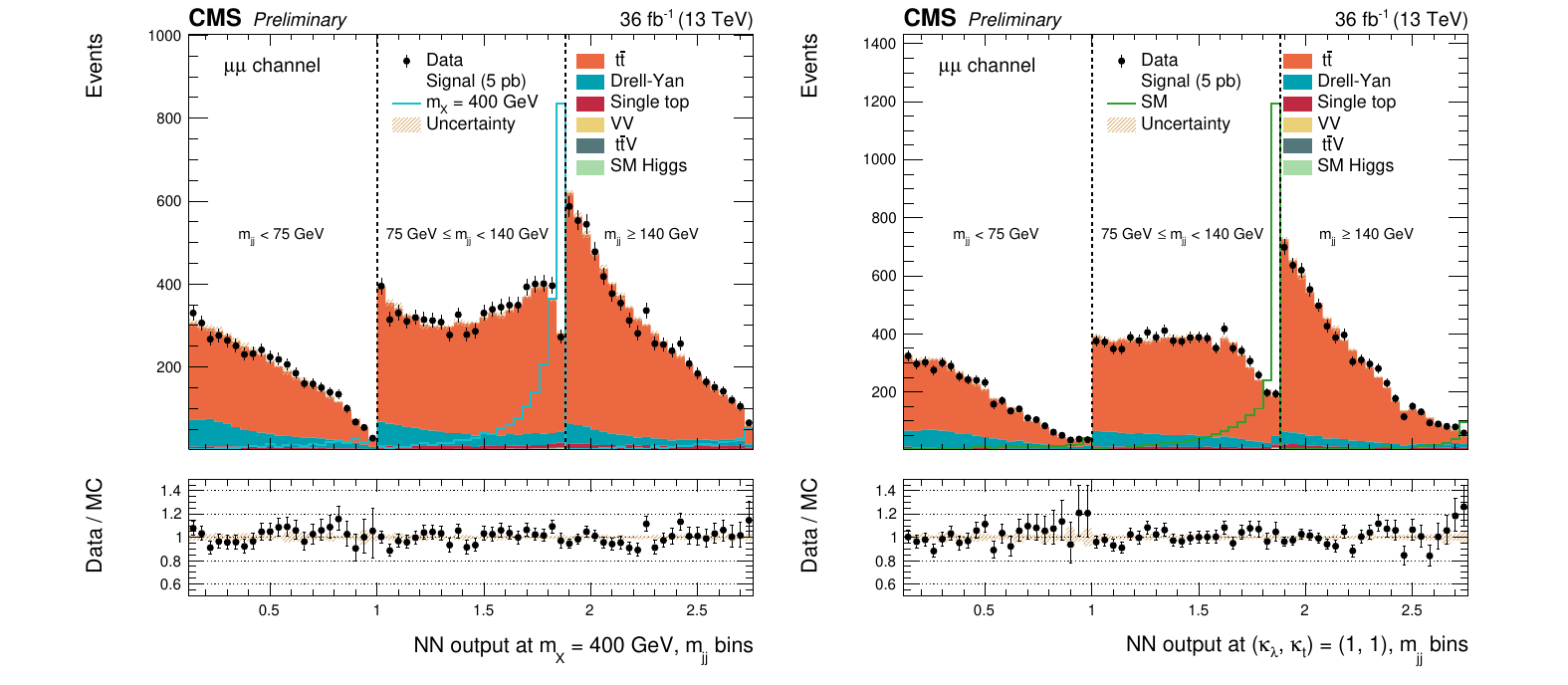}
\caption{H($\mathrm{b\overline{b}}$)H($\ell\nu\ell\nu$)~\cite{CMS:2017ums}. Shown are the DNN output distribution for data and simulated events in three different $m_{jj}$ regions, for the $\mu^+\mu^-$ channel: $m_{jj}<$ 75 GeV, $75<m_{jj}<$ 140 GeV, and $m_{jj}>$ 140 GeV. The resonant DNN output (left) evaluated at $m_X= 400$ GeV and the non-resonant DNN output (right) evaluated at $\kappa_{\lambda}=1$, $\kappa_{t}=$1.}
\label{fig:bblnulnu}
\end{figure}

In this analysis backgrounds are estimated from simulation. Signal discrimination is enhanced by exploiting event kinematics via a parametrized Deep Neural Net (DNN) discriminant, in 3 different dijet mass bins. The output of the DNN on signal and background is shown in Figure \ref{fig:bblnulnu} for both the resonant (left) and non-resonant (right) analyses.

\begin{figure}[htb]
\centering
\includegraphics[width=.7\textwidth]{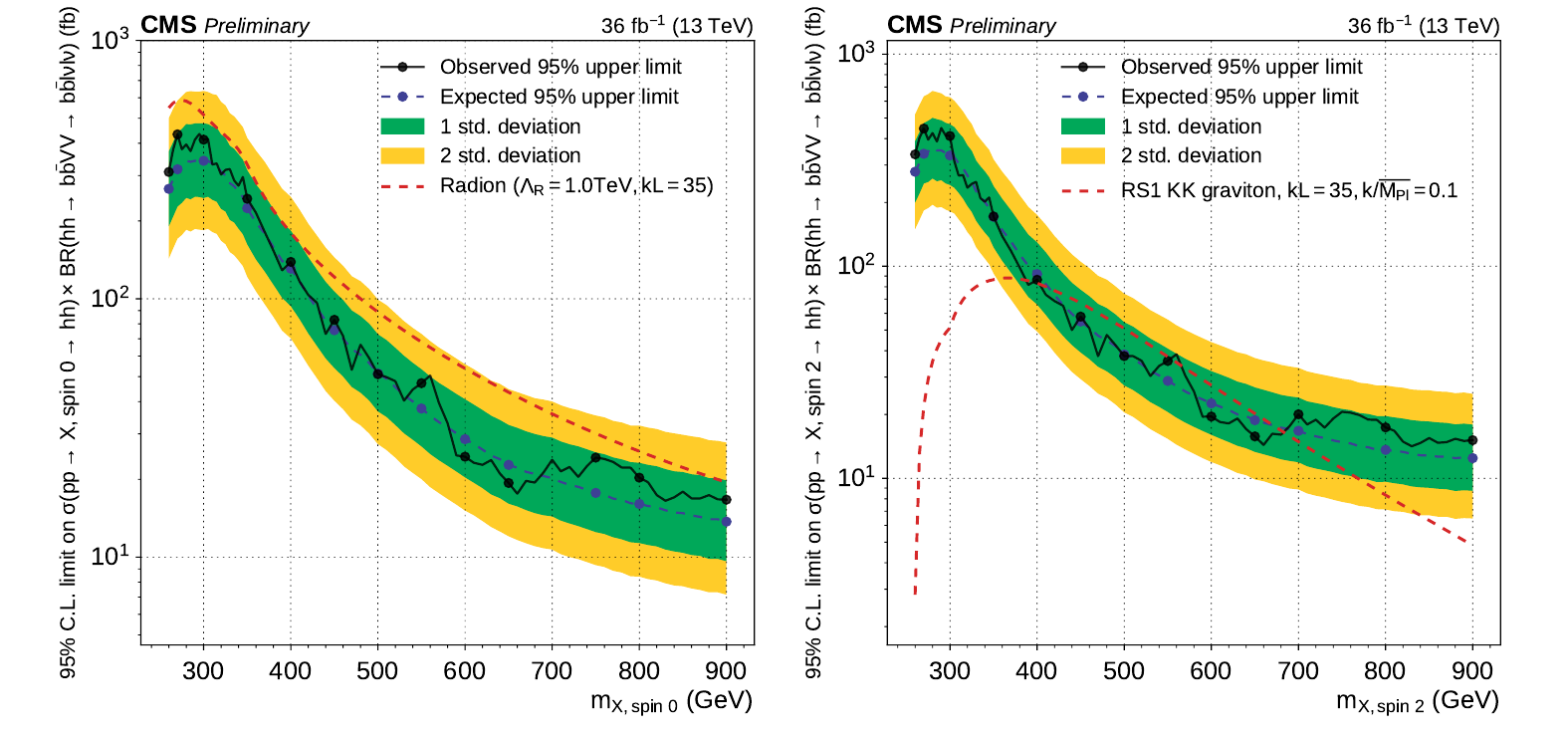}
\caption{Expected and observed 95\% CL upper limits on resonant Higgs pair production cross section times branching ratio for H($\mathrm{b\overline{b}}$)H($\ell\nu\ell\nu$)~\cite{CMS:2017ums} as a function of $m_X$. These limits combine the $e^+e^-$, $\mu^+\mu^-$, and $e\pm\mu\mp$ channels, for spin-0 (left) and spin-2 (right) hypotheses.}
\label{fig:bblnulnu_res}
\end{figure}

In the resonant analysis, results are interpreted in both the spin-0 and spin-2 Radion BSM interpretations. 95\% CL limits on the di-Higgs boson production cross section times branching fraction can be seen in Figure \ref{fig:bblnulnu_res}.

\begin{figure}[htb]
\centering
\includegraphics[width=.66\textwidth]{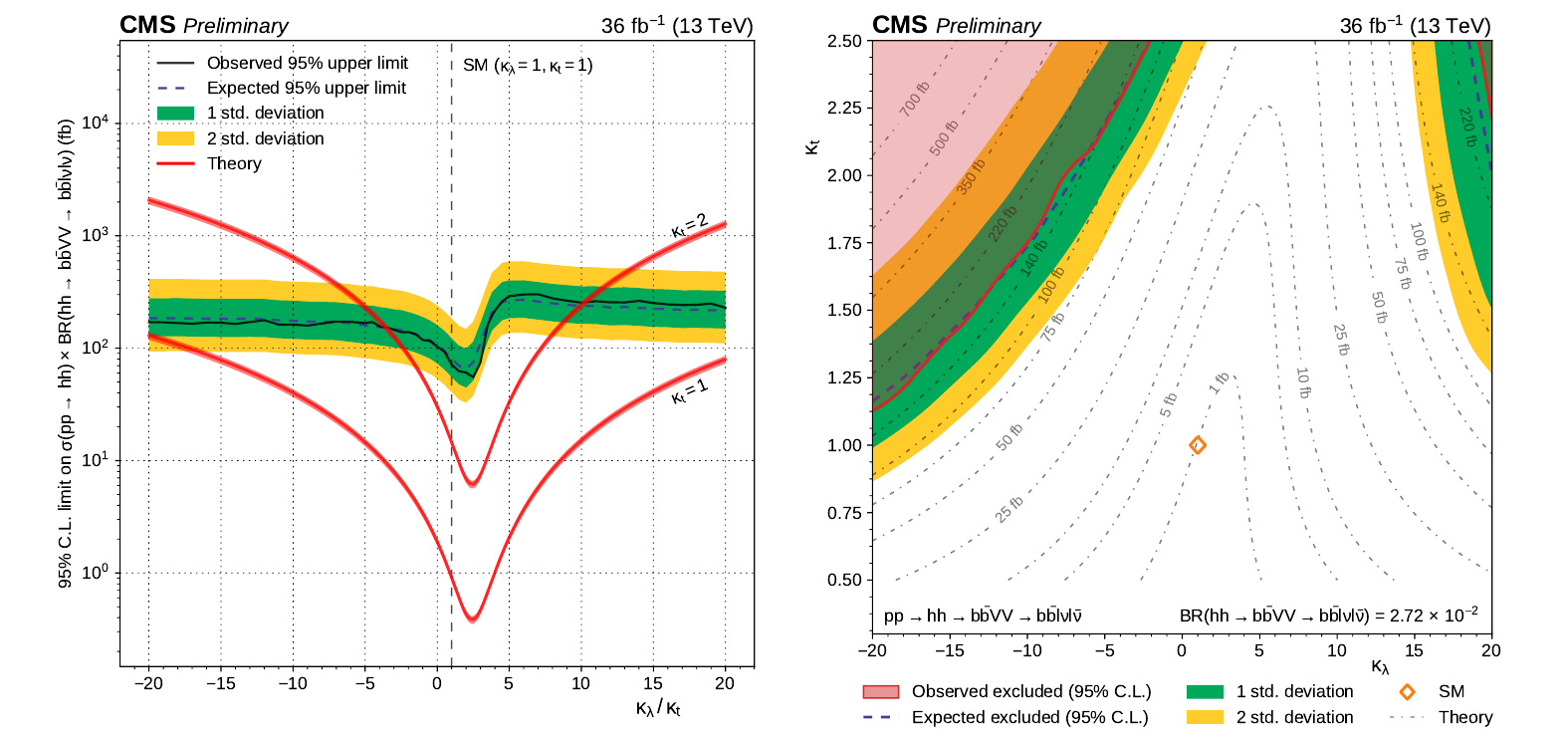}
\caption{Expected and observed 95\% CL upper limits on non-resonant Higgs pair production cross section times branching ratio for H($\mathrm{b\overline{b}}$)H($\ell\nu\ell\nu$)~\cite{CMS:2017ums} as a function of $\kappa_\lambda/\kappa_t$ (left). Observed (red region) and expected (dashed blue line) exclusion of the BSM parameter space at 95\% CL in the $\kappa_\lambda$ vs. $\kappa_t$ plane (right).These limits combine the $e^+e^-$, $\mu^+\mu^-$, and $e\pm\mu\mp$ channels.}
\label{fig:bblnulnu_nonres}
\end{figure}

Non-resonant production is interpreted as a function of the anomalous couplings $\kappa_\lambda$ and $\kappa_t$. Figure \ref{fig:bblnulnu_nonres} shows 95\% CL limits on production time branching fraction as a function of $\kappa_\lambda/\kappa_t$ (left) and in the $\kappa_\lambda$ vs. $\kappa_t$ plane (right). In the SM interpretation, this corresponds to observed (expected) limits of 72 (81) fb.

\subsection[H(bb)H(tau tau)]{H($\mathrm{b\overline{b}}$)H($\tau\tau$)}

\begin{figure}[htb]
\centering
\includegraphics[width=0.3\textwidth]{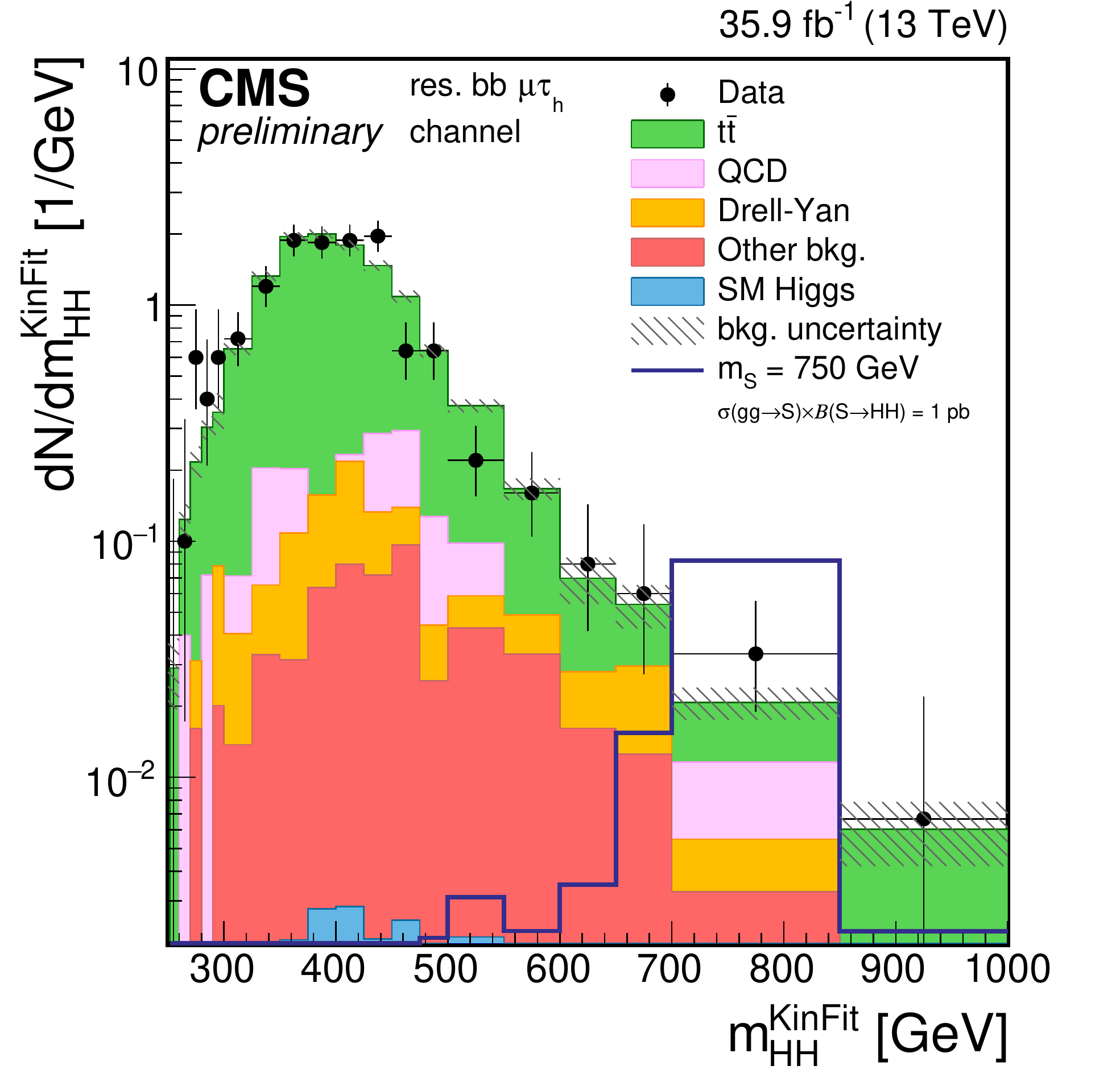}
\includegraphics[width=0.3\textwidth]{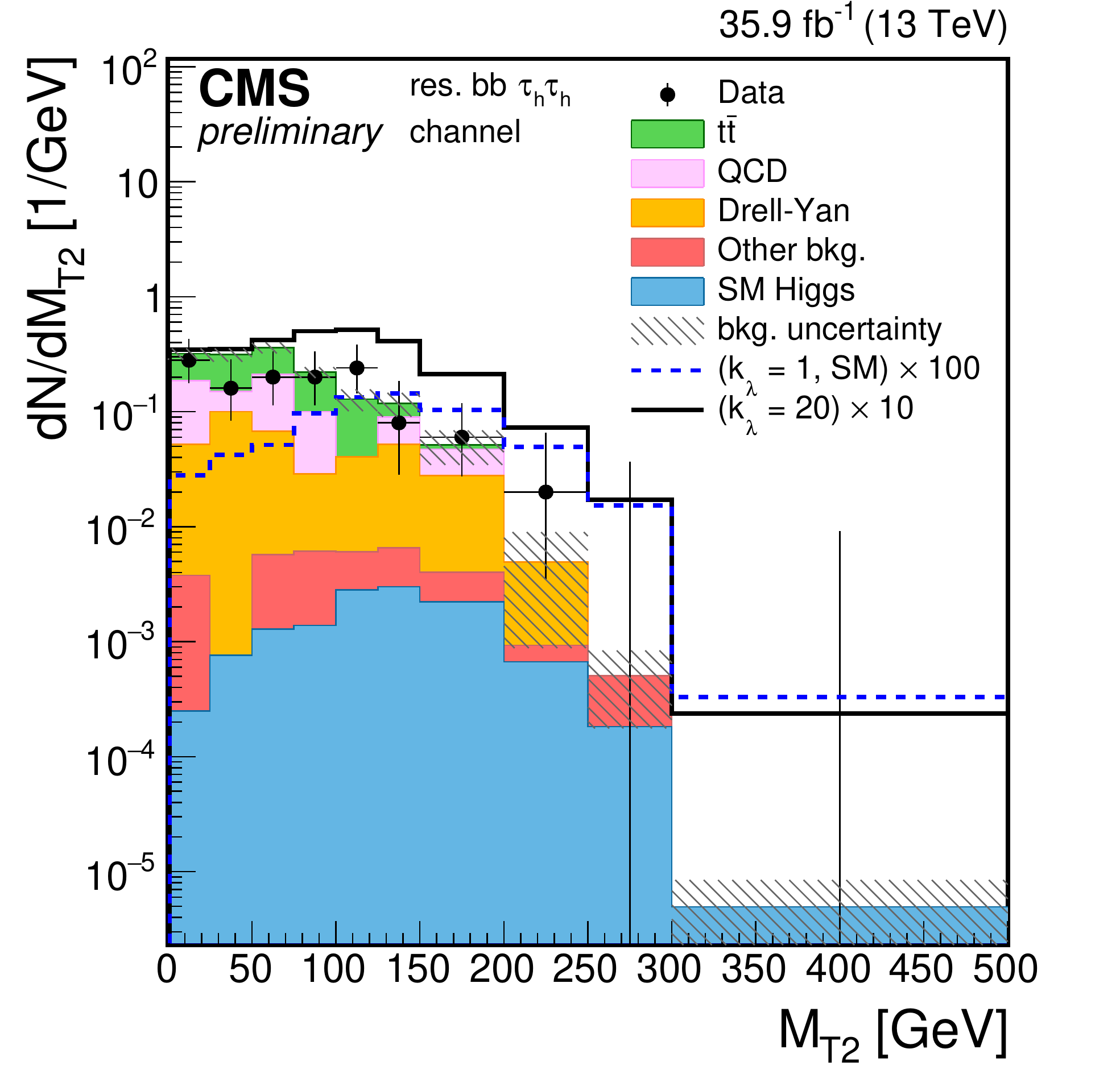}
\caption{H($\mathrm{b\overline{b}}$)H($\tau\tau$) channel~\cite{CMS:bbtt}. Left: Distribution of the $m_{\mathrm{HH}}^{\mathrm{KinFit}}$ variable for events observed in the resolved 2b signal region of the $\tau_{h}\tau_{\mu}$ final state, in the high-mass region. Right: Distribution of the $m_{\mathrm{T2}}$ variable for events observed in the resolved 2b signal region of the $\tau_{h}\tau_{\mu}$ final state.}
\label{fig:bbtt}
\end{figure}

Resonant and non-resonant searches are performed in the H($\mathrm{b\overline{b}}$)H($\tau\tau$) channel using 35.9 fb$^{-1}$~\cite{CMS:bbtt}. The analysis is split into 3 distinct ditau decay channels, H$\rightarrow\tau_{h}\tau_{h}$/$\tau_{h}\tau_{e}$/$\tau_{h}\tau_{\mu}$, and further into 3 categories: 2 b tags, 1 b tag, and a boosted high-mass category. For all categories the dijet and ditau invariant masses are constrained to be consistent with the Higgs boson mass. A boosted decision tree (BDT) discriminant is used to reject the dominant t$\overline{\mathrm{t}}$ background in the $\tau_{h}\tau_{\ell}$ channels. Signal discrimination is performed in the resonant analysis using the di-Higgs boson system kinematic fit $m_{\mathrm{HH}}^{\mathrm{KinFit}}$, which increases resonance-mass resolution, and in the non-resonant analysis using the `stransverse' mass $m_{\mathrm{T2}}$, specifically designed to discriminate between signal and background in topologies with multiple sources of missing momentum. Background to data comparisons of $m_{\mathrm{HH}}^{\mathrm{KinFit}}$ and $m_{\mathrm{T2}}$ can be seen in Figure \ref{fig:bbtt} left and right, respectively.

\begin{figure}[htb]
\centering
\includegraphics[width=0.34\textwidth]{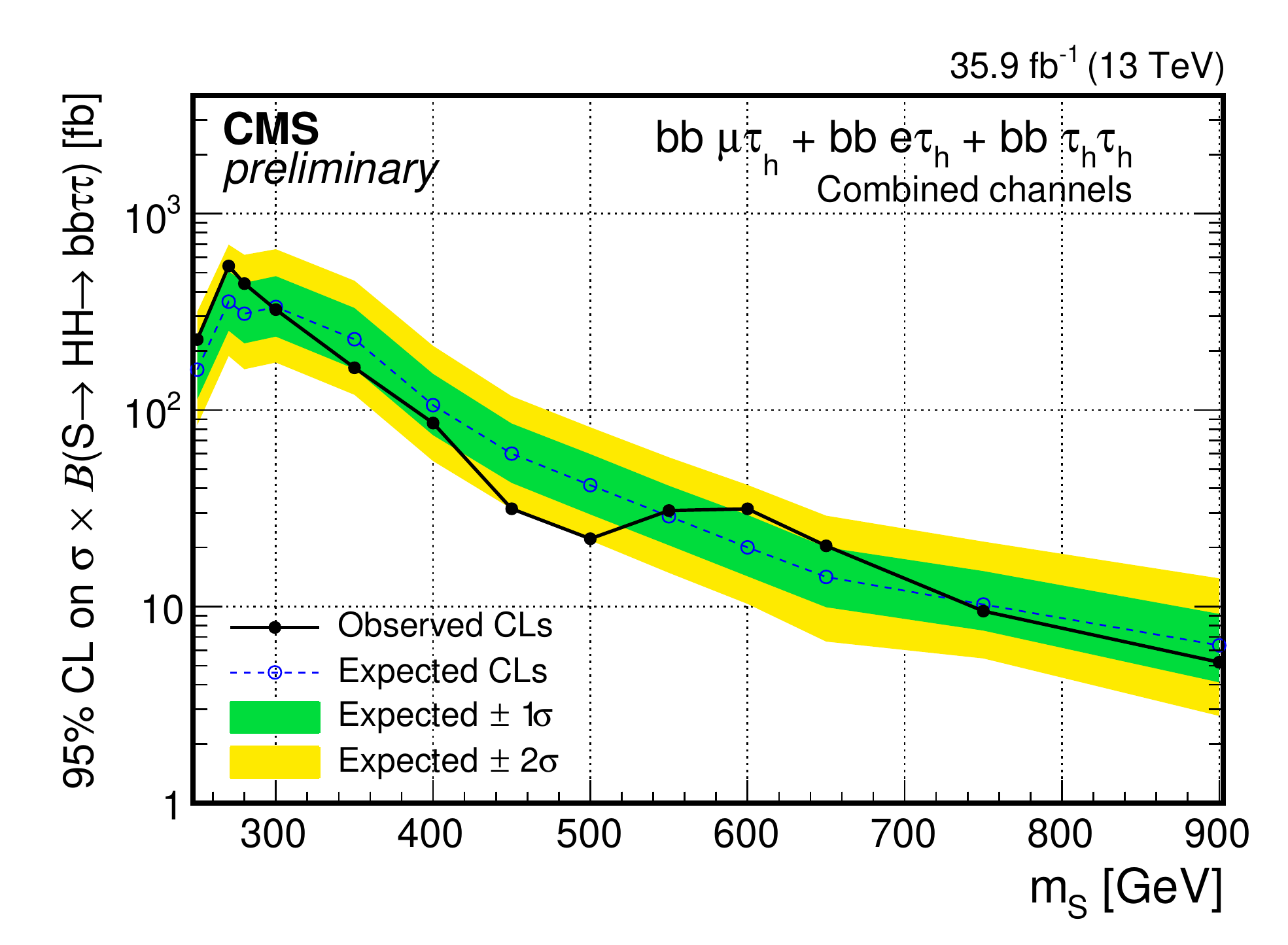}
\includegraphics[width=0.34\textwidth]{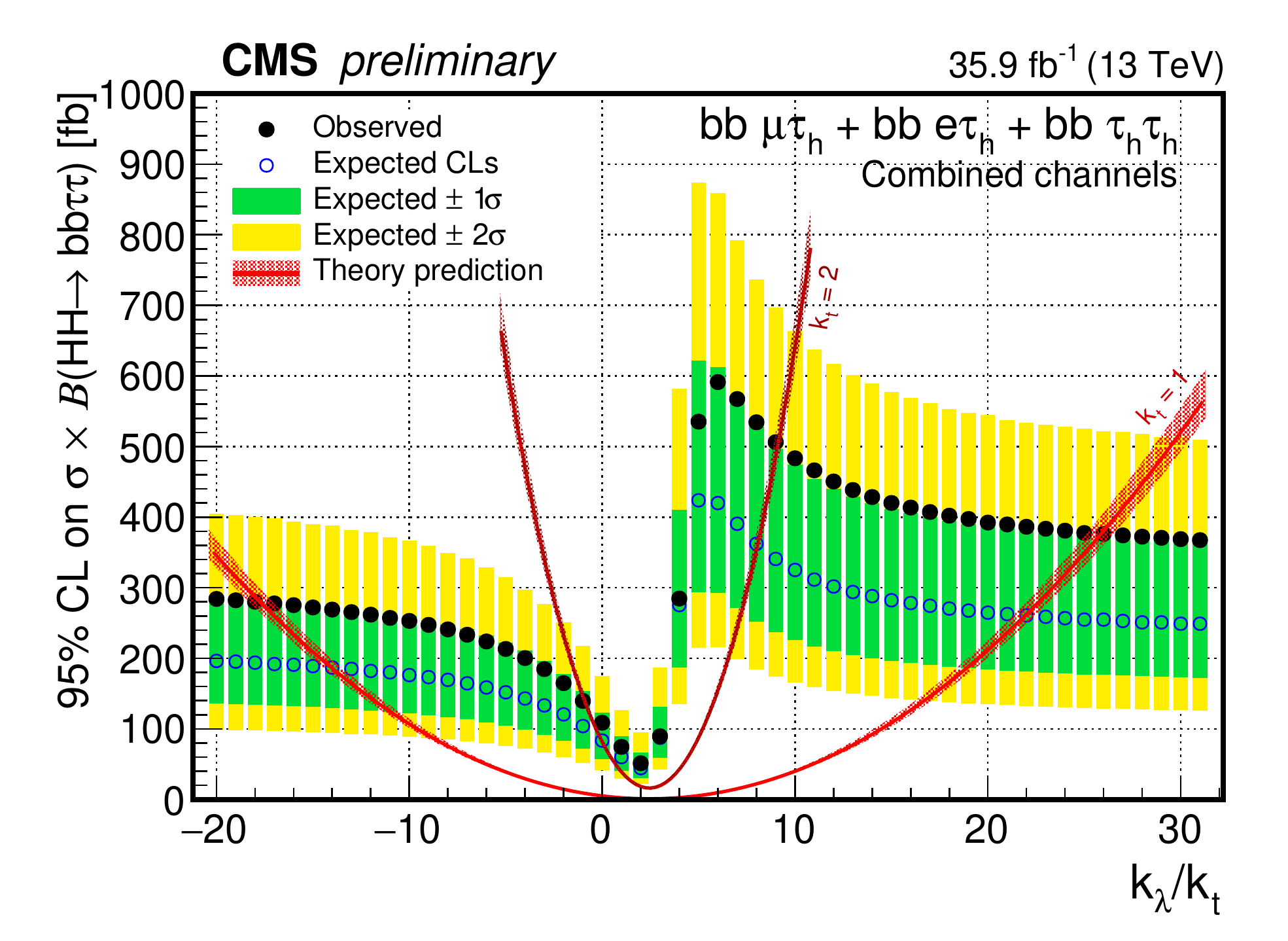}
\caption{Left: Expected and observed 95\% CL upper limits on resonant Higgs pair production cross section times branching ratio as a function of resonance mass $m_S$ for H($\mathrm{b\overline{b}}$)H($\tau\tau$)~\cite{CMS:bbtt}. Right: Observed and expected 95\% CL upper limits on cross section times branching fraction as a function of $\kappa_\lambda/\kappa_t$. }
\label{fig:bbttResults}
\end{figure}

All categories are combined to set limits on di-Higgs boson production cross section times branching fraction. Figure \ref{fig:bbttResults} left (right) shows the resonant (non-resonant) results. In the SM non-resonant hypothesis, this corresponds to observed (expected) limits of 28 (25) times the SM prediction.

\section{Summary of Results and Conclusions}

\begin{figure}[htb]
\centering
\includegraphics[width=0.4\textwidth]{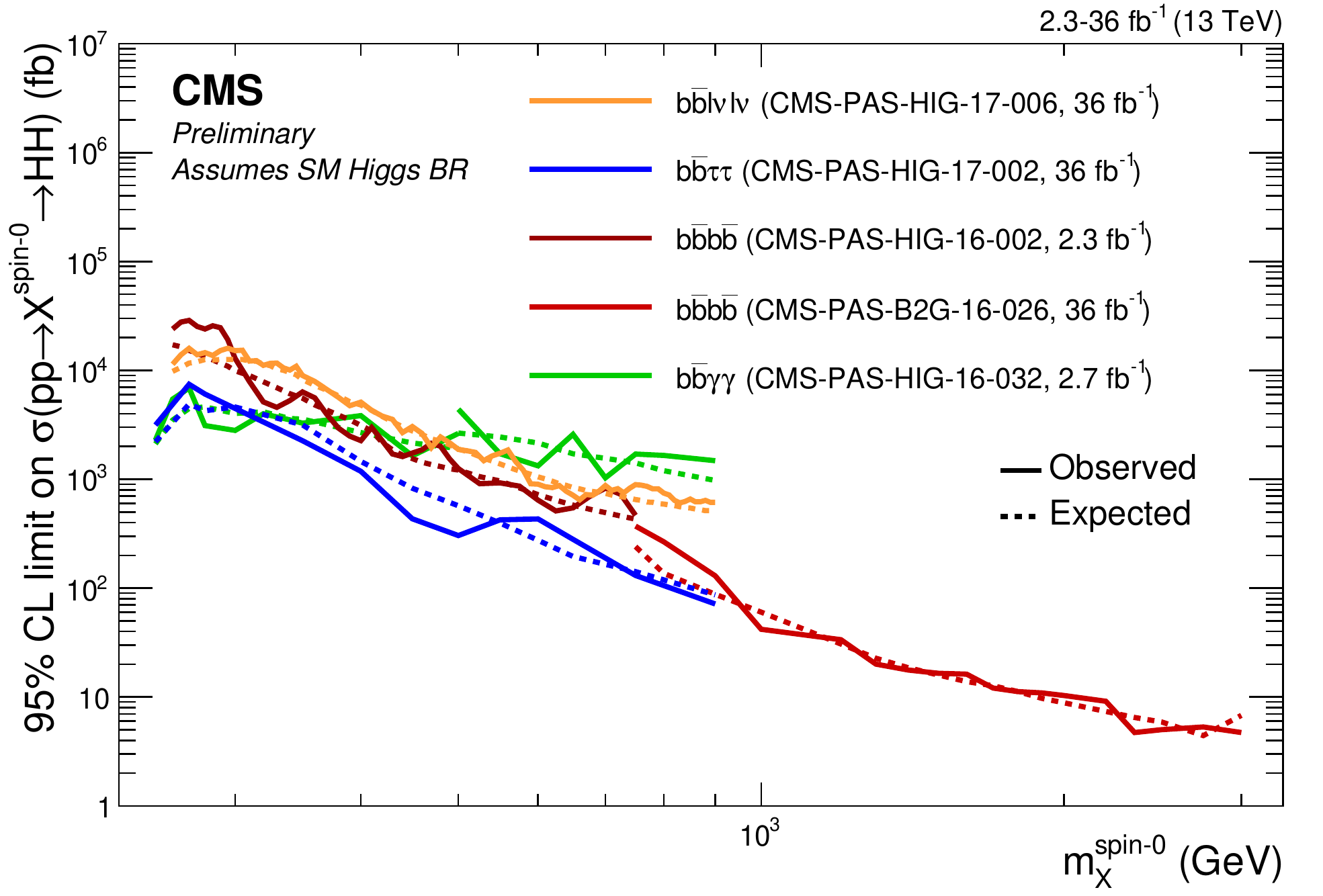}
\includegraphics[width=0.4\textwidth]{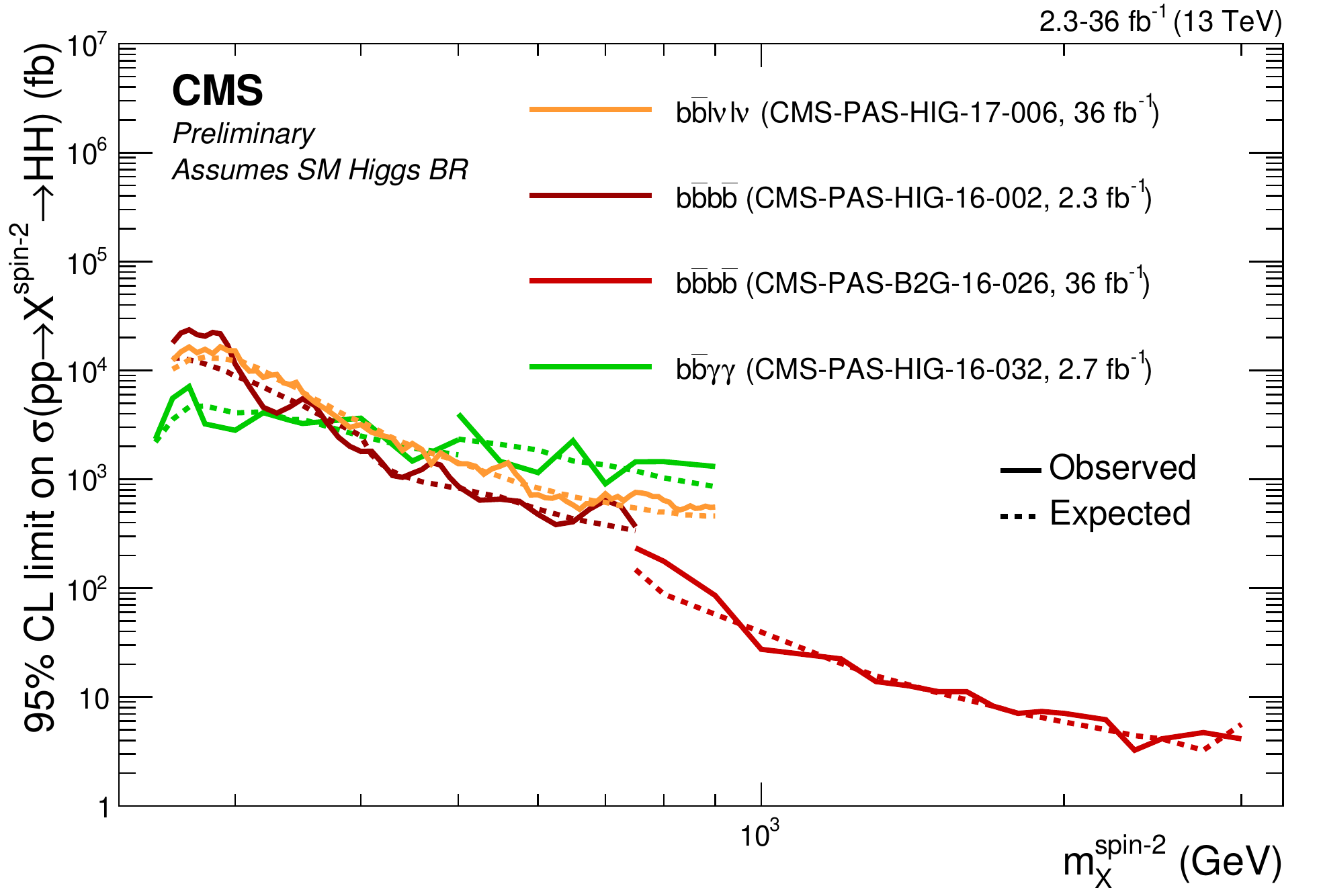}
\caption{Observed and expected 95\% CL upper limits on the product of cross section and the branching fraction $\sigma(\mathrm{gg}\rightarrow X)\times BR(X\rightarrow\mathrm{HH})$ obtained by different analyses assuming spin-0 (left) and spin-2 (right) hypothesis in an extended mass range beyond 1 TeV .}

\label{fig:comb}
\end{figure}

A summary of results of all the CMS di-Higgs boson resonant analyses is shown in Figure \ref{fig:comb}~\cite{CMS:2017gxe}. The complementarity of different analyses, along with the similar sensitivities in the low mass region highlight the necessity of multiple channels, and the need for an eventual combination of the results from all the channels. Results of all the CMS di-Higgs boson non-resonant analyses are summarized in Table \ref{tab:summaryResults}.

Searches for HH production give both insight into the nature of electroweak symmetry breaking and searches for new physics. CMS has a suite of HH searches which are disjoint and complementary across a large mass range. So far no sign has been seen of SM or BSM HH production, with current best limits at 28 times the SM cross section. This is expected to improve as the remaining analyses update to the full 2016 LHC dataset. As the LHC transitions from the current energy-scaling regime to the luminosity-scaling era of the HL-LHC, SM and BSM di-Higgs boson models will be tested. Improving analyses, adding new final states and combining results will be crucial to the continues success of these searches.

\begin{table}[htb]
\begin{center}
\begin{small}
\begin{tabular}{l|cc}  
Analysis &  Observed (Expected) $\sigma / \sigma_{SM}$ limits &  Integrated luminosity ($\mathrm{fb}^{-1}$)  \\ \hline
H($\mathrm{b\overline{b}}$)H($\tau\tau$) & 28 (25) & 35.9 \\
H($\mathrm{b\overline{b}}$)H($\ell\nu\ell\nu$) & 79 (89)  & 35.9 \\
H($\mathrm{b\overline{b}}$)H($\gamma\gamma$) & 91 (90) & 2.7\\
H($\mathrm{b\overline{b}}$)H($\mathrm{b\overline{b}}$) & 342 (308) & 2.3-2.7\\
\hline
\end{tabular}
\caption{13TeV di-Higgs boson non-resonant observed (expected) $\sigma / \sigma_{SM}$ 95\% CL limits}
\label{tab:summaryResults}
\end{small}
\end{center}
\end{table}

\Acknowledgements
I am grateful to the \raisebox{-3.5mm}[0pt][0pt]{\includegraphics[width=0.06\textwidth]{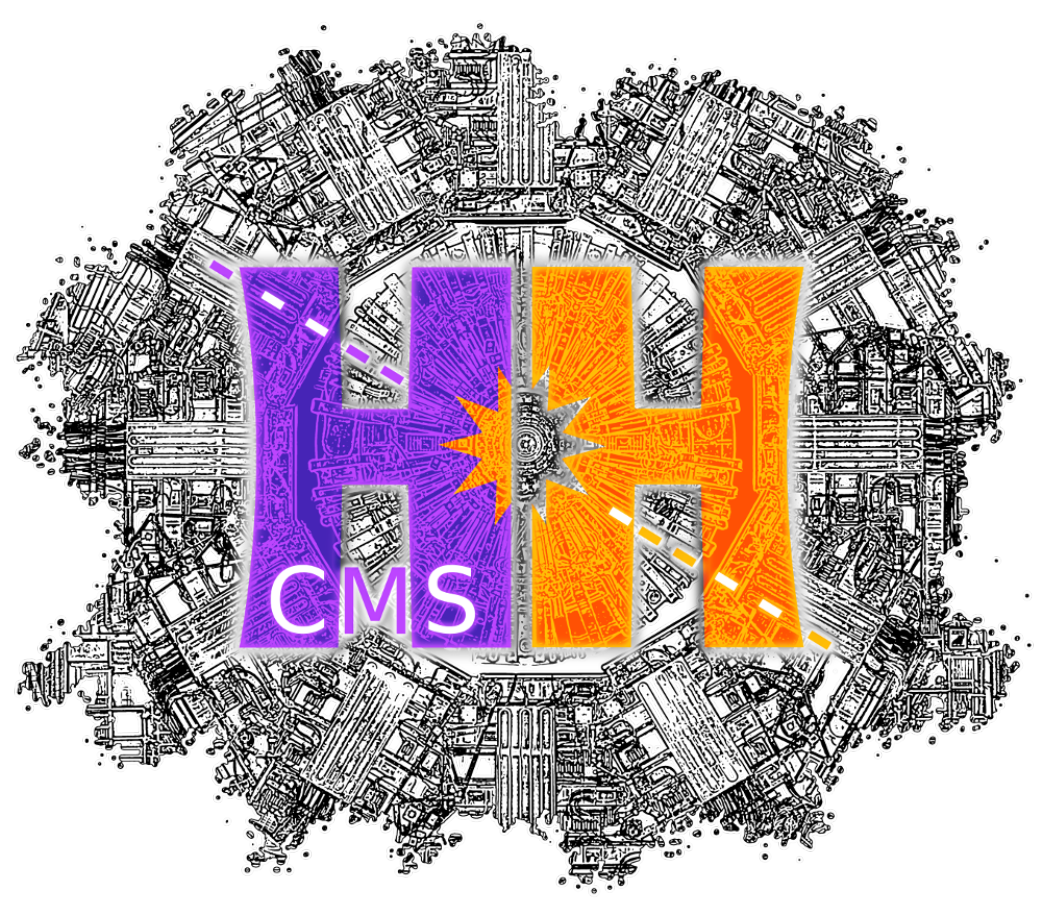}} working group for fruitful discussions.


\begin{thebibliography}{99}


\bibitem{Chatrchyan:2008aa} 
  S.~Chatrchyan {\it et al.} [CMS Collaboration],
  JINST {\bf 3}, S08004 (2008).
  doi:10.1088/1748-0221/3/08/S08004
  
\bibitem{CMS:2016tlj} 
  CMS Collaboration [CMS Collaboration],
  CMS-PAS-HIG-16-002.
  
\bibitem{CMS:2017gxe} 
  CMS Collaboration [CMS Collaboration],
  CMS-PAS-B2G-16-026.

\bibitem{CMS:2016foy} 
  CMS Collaboration [CMS Collaboration],
  CMS-PAS-HIG-16-026.

\bibitem{CMS:2016vpz} 
  CMS Collaboration [CMS Collaboration],
  CMS-PAS-HIG-16-032.
  
\bibitem{CMS:2017ums} 
  CMS Collaboration [CMS Collaboration],
  CMS-PAS-HIG-17-006.
  
  
\bibitem{CMS:bbtt} 
  CMS Collaboration [CMS Collaboration],
  CMS-HIG-17-002, Submitted to Phys. Lett. B.

\end{thebibliography}
\end{document}